\documentclass[smallcondensed]{svjour3}

\usepackage{amssymb,amsmath,graphicx,multirow,url}
\usepackage{pifont,geometry,fleqn,txfonts,hyperref}
\usepackage[numbers]{natbib}

\usepackage[utf8]{inputenc}

\newcommand{\dquotes}[1]{``#1''}   
\newcommand{\squotes}[1]{`#1'}     

\begin{document}

\title{Predicting IR Personalization Performance using Pre-retrieval Query Predictors}

\author{Eduardo Vicente-L\'opez \and Luis M. de Campos \and Juan M. Fern\'andez-Luna \and Juan F. Huete}

\institute{Eduardo Vicente-L\'opez \and Luis M. de Campos* \and Juan M. Fern\'andez-Luna \and Juan F. Huete
	\at Departamento de Ciencias de la Computaci\'on e Inteligencia Artificial, E.T.S.I.I.T., CITIC-UGR, Universidad de Granada, 18071-Granada, Spain \\
	\email{evicente,lci,jmfluna,jhg@decsai.ugr.es} -- *Tel.: +34 958243199; fax: +34 958243317}

\date{Received: date / Accepted: date}

\maketitle

\begin{abstract}
Personalization generally improves the performance of queries but in a few cases it may also harms it. If we are able to predict and therefore to disable personalization for those situations, the overall performance will be higher and users will be more satisfied with personalized systems. We use some state-of-the-art pre-retrieval query performance predictors and propose some others including the user profile information for the previous purpose. We study the correlations among these predictors and the difference between the personalized and the original queries. We also use classification and regression techniques to improve the results and finally reach a bit more than one third of the maximum ideal performance. We think this is a good starting point within this research line, which certainly needs more effort and improvements. 

\end{abstract}

\section{Introduction}
\label{sec:intro}

Digital information is growing exponentially in recent years, which makes more difficult for Information Retrieval Systems (IRSs) to provide truly relevant results. Additionally, traditional IRSs retrieve the same list of results for a given query independently of who submitted it, although the relevance judgments for a given query differ greatly for different users \cite{teevan2010potential}. This is known as the \textit{\squotes{one size fits all}} problem.  Modern IRSs must incorporate new features to face the two previous problems, trying to keep improving their performance and, at the same time, better satisfying the specific user information needs.

Personalization \cite{steichen2012comparative,ghorab2013personalised} is one of these new features. Personalized IRSs consider some knowledge about the user within the retrieval process to try to adapt, as best as possible, the retrieved results to that specific user. Users are generally more satisfied with personalized IRSs, but this is not always the case.

There is a high number of articles \cite{liu2004personalized,dou2007large,micarelli2007personalized,zhou2010evaluating,decampos2014using} that show different ways of integrating personalization within different IRSs and their corresponding performance improvement. But there is one feature they generally do not mention; personalization improves the IRS performance on average, actually for most queries, but it also deteriorates the performance of some queries \cite{teevan2008personalize,dou2007large}. Furthermore, a query could offer good personalized performance results for a given user but not for another one, depending on their user profiles. Therefore, the next step would be to be able to predict this performance in order to apply it or not for each individual query and user.

The personalization performance prediction problem has many similarities with the classical IR query performance prediction \cite{carmel2006makes}, where the quality of the retrieved results for some queries may be poor although the corresponding IRS performs well on average. These poor performance queries are called \textit{difficult} queries, and they should be identified to be handled properly. A difficult query is one that has so many possible answers that it is difficult to choose the correct or the most appropriate one. Very specific or focused queries are easy to be answered, so they are not difficult queries. However, an example of a difficult query is an ambiguous query.

Intuitively, it is easy to see that difficult queries will benefit from personalization, as personalization narrows the query focus on the user and helps disambiguation, among other things. Therefore, the hypothesis is that the prediction of difficult queries and personalization performance will be correlated to some degree.

There are so many articles addressing the query difficulty prediction problem \cite{carmel2006makes,hauff2008improved,he2008using,carmel2010estimating,shtok2012predicting,dan2016measuring}. They try to predict the query performance to make use of adaptive retrieval components in order to outperform the IRS performance. Query difficulty predictors may be classified into two categories: \textit{pre-retrieval} predictors (before the retrieval step), which make use of the information about the query and the document collection at indexing time, and \textit{post-retrieval} predictors (after the retrieval step), which make use of the retrieved results and user interaction behaviour for the query. 

We have only focused on pre-retrieval predictors mainly because of two reasons. The first reason is that we want our findings to be usable in real production IRSs, so time response is important. Pre-retrieval predictors are much faster to be calculated, since they only use information available at indexing time, in contrast to post-retrieval predictors which use different characteristics of the retrieved results (mostly in web environments). The second reason is that our experimental environment is at a much lower scale than usual web IRSs, so we do not have enough information about the query history characteristics such as the number of times issued, the click entropy (variability of results that people click on), etc.

Our final objective is to discern whether to apply personalization or not for a given query and user profile prior to the retrieval process. We have addressed this personalization performance prediction problem by using a comprehensive set of state-of-the-art pre-retrieval query difficulty predictors, plus some others proposed by us. We have correlated their values with those of the difference between the performance of the  personalized and the original query. 

Although some of these predictors are correlated with the personalization performance, the correlations are not very high. For that reason, in order to take the final decision and to try to improve the prediction results we have joined the potential benefit of each of the individual pre-retrieval predictors using classification and regression techniques. Our results show that we are able to reach about one third of the ideal performance (which would be to be able to identify all queries where personalization obtains worse results than the original query, and therefore do not apply it to that query and user profile). By disabling personalization for those cases, we get slightly better IRS performance than the strategy of applying personalization to all queries and users.

As far as we know, there is no other article which has done such a comprehensive study and tested their prediction models, giving the actual improved performance of their predictions against always applying personalization.

The remainder of the article is organized as follows: Section \ref{sec:relatedWork} reviews the related work; Section \ref{sec:predictors} describes the used pre-retrieval predictors to try to estimate the personalization performance and why we have used classification and regression techniques with these predictors. Section \ref{sec:results} shows the evaluation environment and the results obtained; and finally Section \ref{sec:conclusionsFuture} outlines the general conclusions of the article and proposals for future research.

\section{Related Work}
\label{sec:relatedWork}

Our final objective is to predict whether to use or not personalization for a given query and user. As described in the introduction this problem is related with the problem of identifying \textit{difficult} queries, based on the assumption that those queries will benefit much more from personalization. Accordingly, the articles in the literature use query difficulty predictors to accomplish the personalization decision problem. Hence, we start reviewing papers about query difficulty predictor to finish with those works using them for our personalization decision problem.

\textbf{\textit{Query difficulty predictors}:} They are also known as query performance predictors (QPP). There are many articles which try to find features to serve as indicators to identify the difficulty or performance of the query. The predictors are classified into two different categories: pre-retrieval and post-retrieval predictors.

\textit{Pre-retrieval predictors}: they try to predict the query difficulty only based on some characteristics of the query and the document collection, without the use of the retrieved results for that query. Since they use some statistics calculated at indexing time they can be computed very fast, which is very important if we want to use them in production environments. They make use of query features, such as the number or average length of query terms, or document collection features such as the inverse document frequency statistics.

There are many articles evaluating the performance of pre-retrieval predictors but each one using a different evaluation framework, which makes their findings not comparable. The broader study in this sense was carried out in \cite{hauff2010predicting}.

These predictors can be classified into \textit{linguistic} and \textit{statistical} methods \cite{carmel2010estimating}. Linguistic approaches use natural language processing and external semantic resources to search for query ambiguity and polysemy. Statistical approaches check deviations in the distribution of the query terms frequency within the corpus of documents.

Linguistic approaches exhibit a poor query performance prediction as shown by different articles. In \cite{mothe2005linguistic} the authors extract a set of 16 linguistic features of the query with the help of some linguistic tools, such as some syntactical parsers, morphological analyzers and polysemy using WordNet. Most of these linguistic features do not correlate well with system performance. Similar results are presented in \cite{hauff2010predicting} where semantic distances between query terms are calculated using the WordNet taxonomy.

Statistical predictors are classified in turn into four categories: specificity of the query, similarity between the query and the collection, coherence of the query term distribution and finally query term relatedness. A comprehensive review about these categories can be found at \cite{carmel2010estimating}. We present here a brief overview of them.

The \textit{specificity} predictors measure the query terms distribution over the document collection. Thus, the bigger the specificity of a query the easiest will be to answer it and it will also show a better retrieval performance \cite{he2004inferring}. They exploit query terms statistics such as the \textit{idf} or \textit{ictf} \cite{zhao2008effective}. Considering the \textit{similarity} predictors, the bigger the similarity between the query and the corpus the easier will be to answer the query due to the high number of potential relevant documents. In \cite{zhao2008effective} the vector-space query similarity to the collection is calculated considering all the corpus of documents as an unique large document. An example of the \textit{coherency} approach is represented by \cite{he2008using}, which measures the inter-similarity of documents containing the query terms. A much less computationally expensive approach is presented in \cite{zhao2008effective} by measuring the variance of the terms weights over the collection documents containing the term. Each term weight is determined by the IRS weighting scheme, e.g., the widely used \textit{tf-idf}. And finally, the \textit{term relatedness} predictors, which are based on the assumption that highly related query terms lead to well formed queries easy to be answered \cite{hauff2008survey}. 

\textit{Post-retrieval predictors}: these predictors analyze the top ranked results retrieved by a query. Therefore, they are much more complex and time consuming than their counterpart pre-retrieval predictors. Additionally, since they make use of the IRS list of retrieved results, these predictors heavily depend on the retrieval model being used. However, as they are more complex and make use of extra information with respect to the pre-retrieval predictors, as for example the diversity of results or the user interaction with those results, they usually offer a slightly higher prediction quality.

These predictors can be classified into four main categories: \textit{clarity} based methods, \textit{robustness} based methods, \textit{score distribution} based methods, and \textit{user behaviour} based methods.

The \textit{clarity} based methods measure the focus or coherence of the list of results with respect to the corpus. For a good performance query we expect to find a common vocabulary or language between the query and its results. Specifically, if the language of the query results is significantly different of the language of the entire corpus, it means that the query is focused and good performance results are expected. The first definition of clarity predictors was given in \cite{cronen2002predicting} still being considered as the state-of-the-art. Some other approaches based on clarity has been proposed as for example in \cite{amati2004query}.

The \textit{robustness} based methods measure the robustness of the query results. The higher the robustness of the results the easier the query is. It can be measured with respect to three different components: 1) perturbations in the query, i.e., if small changes on the query lead to large changes in the list of results the query is difficult \cite{zhou2007query}; 2) document perturbations, i.e., if introducing noise the ranking of results does not differ to much means the query is robust \cite{zhou2006ranking}; and 3) perturbations in the retrieval models, i.e., if different retrieval models produce very similar list of results the query is robust and easy to answer \cite{aslam2007query}.

The \textit{score distribution} based methods are a less expensive alternative to the two previous approaches, since they do not analyze the top ranked documents but their retrieval score distribution \cite{zhou2007query}. If low scores are observed in the top ranked results, this is likely because of a difficult query. A more recent approach was introduced in \cite{shtok2012predicting}, where authors predict the query performance by estimating the query-drift as the standard deviation of retrieval scores in the top ranked list of results.

The \textit{user behaviour} based methods consider user interactions with the query results to predict the query difficulty. In fact, in \cite{guo2010predicting} the authors use different sources of evidence from queries, results and user interaction logs to train a regression model. Their findings show how user interactions reflect a strong signal of query results quality, as it is shown by the fact that the top two features by predictive power are average click position of results and average number of clicks.

In general, in \cite{hauff2010comparison} the authors present a study where they check if the query performance predictors are correlated with the relevance values assigned by real users. They show that for query suggestions the ratings are mostly uncorrelated, while at a topic level some predictors are moderately correlated. These findings suggest that the intuitions behind such predictors are not enough representative of how users rate query results. This motivates further research into proposing new predictors which better capture the users perception of relevance.

\textbf{\textit{Personalization performance predictions}:} In contrast to the extensively studied research area of query performance predictors, there are not many articles focused on predicting the personalization performance. In this case the objective is clear, a perfect personalization performance predictor (PPP) is the one that is able to always identify those personalized queries which outperform the corresponding not personalized (original) ones.

In general personalization improves the original query performance, but it may even harms search accuracy under some situations \cite{dou2007large}. The previous article present a large-scale evaluation framework for personalized search based on query logs. It also shows that click-based personalization techniques perform better than the profile-based ones. But, unfortunately this imply to have access to really large IRSs logs, which in the vast majority of cases is not possible. 

The first approach within this specific area is \cite{teevan2008personalize}. They extract some query features, explicit relevance judgments and large-scale log analysis of user behaviour for each query to study the variability in user intent, i.e., what each user finds relevant to the same query by their clicks on the list of results (click entropy). They try to identify queries with the most variability or click entropy across users, which benefit the most from personalization, as different users find relevant different results. This can be seen as a measure of query ambiguity too. They find the click entropy and potential for personalization at 10 \cite{teevan2010potential} as the two most correlated features with implicit measures of query ambiguity. With all this data they build query ambiguity predictive models to identify queries that can benefit from personalization. However, they do not actually compare the performance of personalized and not personalized approaches, testing this way if their model is helpful or not. 

Another article with the same above problem is \cite{chen2010predicting}, where the authors use classification and regression techniques but in a completely different way to us. They try to predict whether some given predictors correctly predict the variability of the relevant retrieved results for different users. They rely on the assumption that the higher the variability the higher the personalization performance potential, but they do not compare the results obtained by an original query and those obtained by the personalized one, as we do.

The approach most similar to our in the literature is \cite{younus2013personalization}. They study the correlations between three pre-retrieval and two post-retrieval query difficulty predictors and the performance of personalization using explicit relevance judgments from 25 real users. One of the post-retrieval predictors present the best correlation, and in general, they find that when standard QPP methods say that a query is difficult, the performance of this query is improved by the use of personalization.

Finally, we can also find the personalization prediction problem applied to another domain as recommender systems \cite{zhang2013personalize}. The authors analyze the ranking of recommendation lists and from a risk management perspective they provide a technique to predict if personalization will be helpful or not. The resulting switching algorithm, which decides to apply personalization or not, outperforms the common recommendation algorithms.

\section{Pre-retrieval Predictors for Personalization Performance}
\label{sec:predictors}

Despite the a priori inferiority of pre-retrieval predictors with respect to post-retrieval predictors, mainly because they have much less information to make their predictions, some experiments reveal a reasonably good performance, even comparable to some much more complex post-retrieval approaches \cite{carmel2010estimating}. As stated in the introduction, we are going to use and focus on the pre-retrieval predictors, mainly because we want our findings to be usable in real production IRSs and because we do not have enough historic information.

Next, we show and explain the basics of the comprehensive set of pre-retrieval predictors we have used in this article. It must be noted that the query terms are stemmed and stopwords are removed before calculating the values of predictors.\\

The first two approaches are two simple linguistic predictors.

The number of terms in the query (\textit{numQT}). This predictor is based on the assumption that the higher the number of terms in the query, the more specific and better explained the query will be.

\begin{equation}
\label{eq:numQT}
numQT = \sum_{t \in Q} 1,
\end{equation}
where \textit{Q} is the query and \textit{t} each term of the query.

The average query length (\textit{avgQL}). This predictor is based on the assumption that longer terms are less common in the corpus being this way more specific.

\begin{equation}
\label{eq:avgQL}
avgQL = (\sum_{t \in Q} t_{l}) / numQT,
\end{equation}
where $t_{l}$ is the number of characters of term \textit{t}. \\

The rest of predictors are based on the collection statistics calculated at indexing time, being most of them first seen in \cite{he2004inferring} and \cite{zhao2008effective}.

$\bullet$ The first set of predictors is based on a well-known IR concept, the inverse document frequency (\textit{idf}). Each document collection term has its own \textit{idf} value. If the \textit{idf} value is high it means that the term rarely occurs in documents, so it is a very specific or selective term. The \textit{idf}-based predictors are the following:

\begin{equation}
\label{eq:idf}
sumIDF = \sum_{t \in Q} \log \frac{N}{f_{t}}, \qquad avgIDF = \frac{sumIDF}{|Q|_{t \in V}}, \qquad maxIDF = max_{t \in Q} \log \frac{N}{f_{t}},
\end{equation}
where $N$ is the total number of documents in the collection, $f_{t}$ is the number of documents containing term $t$, $Q$ is the query, $V$ is the collection vocabulary (unique terms), and $|Q|_{t \in V}$ is the query length for terms in $V$.

\textit{sumIDF} will be biased toward longer queries. Therefore, we normalize \textit{sumIDF} by the query length only considering those terms in $V$, as \textit{avgIDF}. An alternative normalization approach is to choose the term with the maximum IDF score, as \textit{maxIDF}.

$\bullet$ The second set of predictors is based on another well-known IR concept, the inverse collection term frequency (\textit{ictf}). Each document collection term has its own \textit{ictf} value. The assumption is the same as with the \textit{idf}, high values means very specific or selective terms. The \textit{ictf}-based predictors are the following:

\begin{equation}
\label{eq:ictf}
sumICTF = \sum_{t \in Q} \log \frac{|C|}{f_{c,t}}, \qquad avgICTF = \frac{sumICTF}{|Q|_{t \in V}}, \qquad maxICTF = max_{t \in Q} \log \frac{|C|}{f_{c,t}},
\end{equation}
where $|C|$ is the number of all terms in collection \textit{C}, $f_{c,t}$ is the frequency of term \textit{t} in \textit{C}.

The \textit{SCS} (simplified clarity score) predictor \cite{he2004inferring} is based on the post-retrieval \textit{CS} (clarity score) predictor. \textit{SCS} is strongly related to \textit{avgICTF} assuming each term only appears once in the query (a reasonable assumption except for extra large queries, which are not very frequent). In this case the \textit{SCS} is calculated as follows:

\begin{equation}
\label{eq:scs}
SCS = \log \frac{1}{numQT} + avgICTF.
\end{equation}

The predictors in Equations \ref{eq:idf}, \ref{eq:ictf} and \ref{eq:scs} are classified under the \textit{specificity} category of pre-retrieval predictors, i.e. how specific the query is. The assumption is that queries with low values (sum, average or maximum), which are queries with very frequent terms, are difficult to satisfy. In particular, the \textit{SCS} predictor measures the specificity of the query also considering the query length.

$\bullet$ The third set of predictors (\textit{SCQ}) are classified under the \textit{similarity} category of pre-retrieval predictors, i.e. how similar are the query and the collection. The assumption is that queries with low similarity values will be difficult to answer. The \textit{SCQ}-based predictors are the following:

\begin{equation}
\label{eq:scq}
\begin{aligned}
& sumSCQ = \sum_{t \in Q} (1 + \ln(f_{c,t})) \ln(1 + N/f_{t}),  \qquad avgSCQ = \frac{sumSCQ}{|Q|_{t \in V}}, \\
& \qquad \qquad maxSCQ = max_{t \in Q} \left((1 + \ln(f_{c,t})) \ln(1 + N/f_{t})\right),
\end{aligned}
\end{equation}
where all components have been already explained in the previous Equations.

$\bullet$ The fourth set of predictors (\textit{VAR}) are classified under the \textit{coherency} category of pre-retrieval predictors, i.e. they measure the inter-similarity of documents containing the query terms. Concretely, \textit{VAR(t)} measures the variance of the term \textit{t} weights over the collection documents containing it. The term weight depends on the used retrieval model. The assumption is that if the variance of the term distribution over the documents containing \textit{t} is low, the query will be more difficult to answer.

Each query term \textit{t} will have a weight value $w_{d,t}$ if it is present in document \textit{d}. Then, the distribution of \textit{t} across all collection documents containing it can be estimated. We use a simple \textit{tf-idf} approach to calculate $w_{d,t}$:

\begin{equation*}
w_{d,t} = (1 + \ln(f_{d,t})) \ln(1 + N/f_{t}),
\end{equation*}
where $f_{d,t}$ is the frequency of \textit{t} in document \textit{d}, with $w_{d,t} = 0$ for query terms not in the collection vocabulary $V$.

To calculate the variance or dispersion we also need the average weight of term \textit{t} over the collection ($\overline{w}_{t}$):

\begin{equation*}
\overline{w}_{t} = \frac{\sum_{d \in D_{t}} w_{d,t}}{f_t},
\end{equation*}
where $D_{t}$ is the set of collection documents containing term \textit{t}, and $f_t$ is its size.

The \textit{VAR}-based predictors are calculated as follows:

\begin{equation}
\label{eq:var}
\begin{aligned}
& sumVAR = \sum_{t \in Q} \sqrt{\frac{1}{f_{t}} \sum_{d \in D_{t}} (w_{d,t} - \overline{w}_{t})^{2}}, \qquad  avgVAR = \frac{sumVAR}{|Q|_{t \in V}}, \\
& \qquad \qquad maxVAR = max_{t \in Q} \sqrt{\frac{1}{f_{t}} \sum_{d \in D_{t}} (w_{d,t} - \overline{w}_{t})^{2}}.
\end{aligned}
\end{equation}

We do not use any pre-retrieval predictor from the \textit{term relatedness} category, e.g. \textit{PMI}, since our retrieval model does not consider term proximity in its retrieval process. As mentioned in \cite{carmel2010estimating}, the queries with strongly related terms will be probably best served by retrieval models which use this term proximity. In \cite{hauff2010predicting} it is shown that \textit{maxVAR} and \textit{maxSCQ} outperform the rest of predictors. Based on this last statement and because the \textit{maxVAR} and \textit{maxSCQ} assumptions are based on different sources of evidence, we also consider valuable to combine both predictors to see if the combination improves the prediction accuracy.

We use the same simple interpolation given in \cite{zhao2008effective} to combine both kinds of predictors as follows:

\begin{equation}
\label{eq:joints}
\begin{aligned}
& joint = \alpha \; maxSCQ + (1-\alpha) sumVAR, \\
& joint2 = \alpha \; maxSCQ + (1-\alpha) maxVAR,
\end{aligned}
\end{equation}
where $\alpha$ is a parameter to determine the importance of each component within the interpolation. We have selected a value of $\alpha = 0.75$, which is within the maximum performance range as the authors of the previous article suggest. Surprisingly, although the best performance predictors according to \cite{hauff2010predicting} are \textit{maxVAR} and \textit{maxSCQ}, the authors in \cite{zhao2008effective} use \textit{maxSCQ} and \textit{sumVAR} in their \textit{joint} predictor. Therefore, we have also proposed a \textit{joint2} predictor using \textit{maxVAR} instead of \textit{sumVAR}.\\

\subsection{Including the profile in predictors}
\label{subsec:predictorsQP}

All the previous seventeen pre-retrieval predictors only use information from the query and the collection. This is because all of them have been designed as predictors for the query difficulty, but not specifically for personalization prediction purposes. When we consider personalization, a third component comes into play, the user profile. There are different ways to collect the user information and to represent it \cite{vicente2016use}. In this article, we are going to work with user profiles represented as a set of weighted keywords.

When we apply personalization, if the query terms are very different from the user profile terms, the information of the user profile will steer the query results to the interests and preferences of the user. However, if the user profile terms are very similar to those in the query, the effect of including the user profile information will not affect too much to the original query results, since the original query was already informative enough and close to the user. Therefore, the bigger the difference between the query and the user profile the higher the impact of personalization. To measure this difference, and as our first proposed predictor considering the user profile, we propose to use a simple cosine similarity measure between the query and the user profile.

The \textit{cosine} predictor is calculated as follows:

\begin{equation}
\label{eq:cosine}
cosineQP = \frac{\sum_{t \in Q\cap Prof} wq_{t} \; wp_{t}}{\sqrt{\sum_{t \in Q} wq_{t}^{2}} \sqrt{\sum_{t \in Prof} wp_{t}^{2}}},
\end{equation}
where $wq_t$ and $wp_t$ are the weights of term $t$ in the query $Q$ and the profile $Prof$, respectively.

The values of \textit{cosineQP} ranges from $0$ meaning totally different terms to $1$ meaning exactly the same terms between the query and the user profile, respectively.\\

The next step is to consider the three components involved in the personalization process at the same time, the query, the user profile and the document collection. One of the most common ways to personalize a query is to simply add a given number of the user profile terms as expansion terms to the original query. Under this case, all the previous used pre-retrieval predictors only considering the query and the document collection could be reused, but in this case using the expanded query instead of their original query.

Hence, we can modify all the previous pre-retrieval predictors (excluding \textit{numQT} and \textit{avgQL}, where the modification has not sense) in Eq.\ref{eq:idf} to Eq.\ref{eq:joints}, which will be denoted by adding \textit{QP} to the end of their names, e.g., \textit{sumIDFQP, avgICTFQP} or \textit{maxSCQQP}.

Another approach to take into account both the query and the profile is to consider two separate queries, the original and the expanded query including the profile terms. We can compute the difference between \textit{avgIDFQP} and \textit{avgIDF}. If this difference is positive, it means that the expanded query is more specific than the original query, hence it is probably more easy to satisfy. Notice that this argument is valid if we use the \textit{avg} version of the predictor but it is not for the \textit{sum} version. We can extend by analogy this reasoning to the other families of predictors, thus proposing the following four predictors:

\begin{equation}
\label{eq:diffOnAvgs}
\begin{aligned}
& profIDF = avgIDFQP - avgIDF, \qquad profICTF = avgICTFQP - avgICTF, \\
& profSCQ = avgSCQQP - avgSCQ, \qquad profVAR = avgVARQP - avgVAR.
\end{aligned}
\end{equation}

All the previous proposed approaches make a total of 37 different pre-retrieval PPP, which conform a quite comprehensive and heterogeneous set of predictors focusing in different components such as the query, document collection or user profile and in different aspects such as the specificity, similarity or coherency of the previous components.

\section{Experimental Environment and Results}
\label{sec:results}

This section shows all the necessary components to perform the evaluation process to check if we can accurately predict the personalization performance and use these predictions to decide whether to personalize or not a given query. It also shows the obtained results and the main conclusions.

\subsection{Experimental framework}
\label{subsec:expFramework}

Our document collection is composed by a set of official documents from the Andalusian Parliament in XML format. More concretely it is composed by 658 committee sessions from the sixth and seventh terms of office (containing 432,575 retrievable structural units). Each committee session is devoted to different areas of interest such as agriculture, education, economy, etc. Each of these documents contains the transcriptions of the speeches of the members of the Parliament discussing about a proposed initiative relative to a given issue in the corresponding committee session.

We use Garnata \cite{decampos2006garnata} as the search engine, which is based on probabilistic graphical models. This structured IRS has been tested and improved at three editions of the INEX workshop \cite{decampos2009new}. From the thirteen personalization strategies proposed in \cite{decampos2014using} we use the so-called HardReranking (HRR) approach, which although not being the one with the best performance, it is the best within the techniques which are more easily implementable by other IRSs.

For the evaluation of the personalized results we still need a set of queries, users and their user profiles and relevance assessments. We have two different sets of the previous components. A first set from a carried out user study \cite{decampos2014using} and a second set from an automatic strategy for personalized IRSs evaluation called ASPIRE \cite{vicente2015automatic}.

\textit{User study}: We have an heterogeneous set of 23 queries formulated by real users of the document collection, which represents a small but trustworthy sample of real user information needs. The user study involved 31 users. Each user submitted one or several of the previous 23 queries to the IRS assuming, among a fixed set of generic profiles, the profile(s) that fits more to him/her. These generic user profiles were automatically learned from the content of the documents in each committee session and represented as sets of weighted terms \cite{vicente2016use}. There are eight different generic user profiles related to administration, agriculture, culture, economy, education, employment, environment and health. When a user evaluates a query under a given profile, a set of relevance assessments is obtained for this user, profile and query. We call the previous set of relevance assessments an \textit{evaluation triplet}. A total number of 126 evaluation triplets were obtained in the user study. For more information about this user study, \cite{decampos2014using} may be checked.

\textit{ASPIRE}: The main problem of user studies is the limited number of the obtained evaluation triplets, due to the enormous effort required to carry out any user study. This problem is solved by ASPIRE. Using this automatic strategy for the evaluation of personalized IRSs we can automatically generate personalized relevance assessments for any given query and profile. It basically considers a query result as relevant if it belongs to the area of interest the user profile represents and it is within the top ranked results, given by a threshold which in our case is equal to 100. Using ASPIRE we have been able to increase the number of queries in order to get more reliable evaluation results. Concretely, for each initiative in the document collection we have used the text within the tag \textit{abstract}, which summarizes in a few words the content of the initiative, as a query to the system. Thus, under ASPIRE we finally have a total of 2602 queries and evaluation triplets with their corresponding sets of relevance assessments.

We will use the Normalized Discounted Cumulative Gain (NDCG) \cite{jarvelin2002cumulated} as the evaluation metric to measure the IRS retrieval performance for a given list of results. This evaluation metric estimates the cumulative relevance gain observed by a user for the top documents in a retrieved list of results. We concretely evaluate this list of results for the top fifty elements.

We next define the improvement of personalization with respect to the original query as the difference between the performances (measured by NDCG) of the personalized and the original query:

\begin{equation}
\label{eq:diffPerso}
diffPerso = NDCG@50_{HRR} - NDCG@50_{Orig}.
\end{equation}

Table \ref{tab:tripletsByProfile} shows the distribution of the evaluation triplets of both the user study and ASPIRE approaches for each user profile. It also shows the number of evaluation triplets where \textit{diffPerso} is positive (personalization outperforms the original query), negative (personalization harms the original query) and equal to zero (neither improvement nor loss in performance).

\begin{table}[htbp]
	\centering
	\caption{Number of evaluation triplets by profile and \textit{diffPerso} for the User Study and ASPIRE.}
	\resizebox{\linewidth}{!}{
		\begin{tabular}{|c|c|cccccccc|} \hline
			& \textit{diffPerso} & administration & agriculture & culture & economy & education & employment & environment & health \\
			\hline
			\multirow{4}{*}{User Study} & + & 10 & 18 & 14 & 10 & 12 & 10 & 8 & 14 \\
			& - & 4 & 2 & 5 & 6 & 2 & 3 & 4 & 2 \\
			& 0s & 0 & 0 & 0 & 1 & 0 & 1 & 0 & 0 \\ \cline{2-10}
			& Total & 14 & 20 & 19 & 17 & 14 & 14 & 12 & 16 \\ \hline \hline
			\multirow{4}{*}{ASPIRE} & + & 292 & 327 & 376 & 137 & 351 & 169 & 231 & 240 \\
			& -      & 76  & 50  & 127 & 103 & 34  & 10  & 18  & 29 \\
			& 0s     & 2   & 7   & 4   & 3   & 7   & 0   & 9   & 0 \\ \cline{2-10}
			& Total & 370 & 384 & 507 & 243 & 392 & 179 & 258 & 269 \\			
			\hline
		\end{tabular}
	}
	\label{tab:tripletsByProfile}
\end{table}

The total number of ASPIRE evaluation triplets is 20.65 times the number of user study triplets. In terms of \textit{diffPerso} distribution for both approaches there are always more positive than negative cases. For a general idea on this distribution, in the User Study the ratio of negative cases with respect to the positive cases ranges approximately from 10 to 60\% with an average of 29\%. In ASPIRE this ratio ranges from 6 to 33\% (with the exception of the \textit{economy} profile with 75\%) and an average of 21\% considering all the profiles. The number of \textit{0s} is negligible for both approaches. With this data we may consider both approaches roughly comparable in terms of \textit{diffPerso} distribution, although not in terms of the total number of evaluation triplets. But precisely that was the objective of using ASPIRE, i.e., to increase the number of evaluation triplets in order to obtain more robust results and conclusions.

\subsection{Correlations between predictors and \textit{diffPerso}}
\label{subsec:correlations}

Remember that our final objective is to discern whether to apply personalization or not for a given query and user profile prior to the search is performed. If the performance of the search applying personalization is higher than the original query performance, the user will be more satisfied with the personalized results. The measure of this user better or worse satisfaction can be approximated by the difference between the personalized and the original query performances (\textit{diffPerso}). If any of the PPP values in section \ref{sec:predictors} highly correlates with \textit{diffPerso}, we could use this given PPP to predict if personalization will benefit or harm the user perceived IRS quality. Thus, we will be able to activate or not the personalization before the search is performed, maximizing the overall IRS performance for the given user.

The correlation tables for the User Study and ASPIRE using all the 37 proposed predictors and the 8 user profiles are presented in the Appendix \ref{app:appendixA}, for space reasons (two tables of almost one page each). However, to summarize this information, Tables \ref{tab:corsUserStudy} and \ref{tab:corsAspire} present the \textit{diffPerso}-predictors average ($\mu$) and maximum ($max$) correlations grouped by profiles for the User Study and ASPIRE, respectively. We have used the Pearson method implemented in the R statistical framework\footnote{\url{https://cran.r-project.org/}} for all the correlations\footnote{We have also used the Spearman and Kendall methods to compute the correlations, obtaining similar results.}.

\begin{table}[htbp]
	\centering
	\caption{\textit{diffPerso}-predictors average ($\mu$) and maximum ($|max|$) correlation values for all profiles by predictor for the User Study.}
	\resizebox{\linewidth}{!}{
		\begin{tabular}{|c|cc|c|cc|c|cc|c|cc|}
			\hline
			\textbf{predictor} & $\mu$ & \textbf{$max$} & \textbf{predictor} & $\mu$ & \textbf{$max$} & \textbf{predictor} & $\mu$ & \textbf{$max$} & \textbf{predictor} & $\mu$ & \textbf{$max$} \\ \hline			
			numQT & -0.125 & -0.572 & avgSCQ & -0.066 & -0.463 & maxIDFQP & -0.101 & -0.372 & maxVARQP & -0.042 & -0.471 \\ 					
			avgQL & 0.023 & 0.41 & maxSCQ & -0.232 & -0.751 & sumICTFQP & -0.085 & -0.619 & jointQP & -0.062 & -0.492 \\ 						
			sumIDF & -0.204 & -0.725 & sumVAR & -0.164 & -0.479 & avgICTFQP & -0.052 & -0.542 & joint2QP & -0.042 & -0.471 \\ 						
			avgIDF & -0.048 & -0.472 & avgVAR & -0.193 & -0.491 & maxICTFQP & -0.192 & -0.535 & profIDF & 0.041 & 0.459 \\ 						
			maxIDF & -0.216 & -0.673 & maxVAR & -0.174 & -0.481 & SCSQP & -0.028 & -0.494 & profICTF & -0.092 & -0.57 \\ 						
			sumICTF & -0.085 & -0.619 & joint & -0.164 & -0.489 & sumSCQQP & -0.223 & -0.755 & profSCQ & 0.06 & 0.451 \\ 						
			avgICTF & 0.085 & 0.572 & joint2 & -0.173 & -0.493 & avgSCQQP & -0.219 & -0.681 & profVAR & 0.149 & 0.482 \\ 						
			maxICTF & -0.126 & -0.483 & cosineQP & -0.254 & -0.755 & maxSCQQP & -0.027 & -0.052 & -- & - & - \\
			SCS & 0.14 & 0.583 & sumIDFQP & -0.204 & -0.725 & sumVARQP & -0.062 & -0.492 & -- & - & - \\
			sumSCQ & -0.223 & -0.755 & avgIDFQP & -0.193 & -0.652 & avgVARQP & -0.038 & -0.509 & -- & - & - \\ \hline				
		\end{tabular}
	}
	\label{tab:corsUserStudy}
\end{table}

\begin{table}[htbp]
	\centering
	\caption{\textit{diffPerso}-predictors average ($\mu$) and maximum ($|max|$) correlation values for all profiles by predictor for ASPIRE.}
	\resizebox{\linewidth}{!}{
		\begin{tabular}{|c|cc|c|cc|c|cc|c|cc|}
			\hline
			\textbf{predictor} & $\mu$ & \textbf{$max$} & \textbf{predictor} & $\mu$ & \textbf{$max$} & \textbf{predictor} & $\mu$ & \textbf{$max$} & \textbf{predictor} & $\mu$ & \textbf{$max$} \\ \hline 
			numQT & -0.016 & -0.315 & avgSCQ & -0.105 & -0.206 & maxIDFQP & 0.072 & 0.122 & maxVARQP & 0.026 & 0.137 \\
			avgQL & -0.093 & -0.203 & maxSCQ & -0.012 & -0.142 & sumICTFQP & -0.002 & -0.229 & jointQP & 0.016 & 0.187 \\
			sumIDF & -0.002 & 0.164 & sumVAR & 0.024 & 0.137 & avgICTFQP & 0.042 & -0.189 & joint2QP & 0.026 & 0.137 \\
			avgIDF & -0.028 & 0.184 & avgVAR & 0.063 & 0.141 & maxICTFQP & 0.081 & 0.158 & profIDF & 0.032 & -0.185 \\
			maxIDF & 0.054 & 0.108 & maxVAR & 0.03 & 0.137 & SCSQP & 0.041 & 0.288 & profICTF & -0.003 & 0.184 \\
			sumICTF & -0.002 & -0.229 & joint & 0.024 & 0.137 & sumSCQQP & -0.031 & -0.304 & profSCQ & 0.11 & 0.203 \\
			avgICTF & 0.008 & -0.186 & joint2 & 0.03 & 0.137 & avgSCQQP & -0.043 & -0.201 & profVAR & -0.068 & -0.132 \\
			maxICTF & 0.074 & 0.141 & cosineQP & -0.279 & -0.376 & maxSCQQP & 0.029 & 0.059 & -- & - & - \\
			SCS & -0.018 & 0.292 & sumIDFQP & -0.002 & 0.164 & sumVARQP & 0.016 & 0.187 & -- & - & -  \\
			sumSCQ & -0.031 & -0.304 & avgIDFQP & 0.017 & -0.173 & avgVARQP & 0.03 & 0.185 & -- & - & - \\ \hline			
		\end{tabular}
	}
	\label{tab:corsAspire}
\end{table}

We can draw the following conclusions from Tables \ref{tab:corsUserStudy} and \ref{tab:corsAspire}: 1) no one predictor has a high enough average correlation to be considered as a good and robust predictor for the personalization performance; 2) average and especially maximum correlations are higher in the user study than with ASPIRE, this may be due to the low number of evaluation triplets for each profile considered in the user study; and 3) \textit{cosineQP} is the best predictor in terms of correlation both on average and maximum values.

According to the previous results, there is a high variability in the correlation values between \textit{diffPerso} and the predictors. These correlations are very dependent on the given predictor and the applied user profile (see Appendix \ref{app:appendixA}). Therefore, although some predictors are good for some profiles, they are not for others. In many cases, even the same predictor gives negative correlation with some profiles and positive with others.

Table \ref{tab:corsBests} shows the ten best predictors based on their average performance across the different user profiles in our experiments. These would be the predictors we would select to predict the personalization performance gain if we would stop our research here.

\begin{table}[htbp]
	\footnotesize
	\centering
	\caption{Ten best \textit{diffPerso}-predictors average correlation values for the User Study and ASPIRE.}
	\begin{tabular}{|cc|cc|}
		\hline
		\multicolumn{2}{|c|}{\textbf{User Study}} & \multicolumn{2}{|c|}{\textbf{ASPIRE}} \\ \hline
		\textbf{predictor} & \textbf{$\mu$} & \textbf{predictor} & \textbf{$\mu$} \\ \hline
		cosineQP & -0.254 & cosineQP & -0.279 \\
		maxSCQ & -0.232 & profSCQ & 0.11 \\
		sumSCQ & -0.223 & avgSCQ & -0.105 \\
		avgSCQQP & -0.219 & avgQL & -0.093 \\		 
		maxIDF & -0.216 & maxICTFQP & 0.081 \\
		sumIDF & -0.204 & maxICTF & 0.074 \\
		avgIDFQP & -0.193 & maxIDFQP & 0.072 \\
		avgVAR & -0.193 & profVAR & -0.068 \\
		maxICTFQP & -0.192 & avgVAR & 0.063 \\
		maxVAR & -0.174 & maxIDF & 0.054 \\ \hline
	\end{tabular}
	\label{tab:corsBests}
\end{table}

As mentioned previously, \textit{cosineQP} is consistently the best predictor, followed by different variations of \textit{SCQ}. However, none of the predictors may be considered as a good personalization performance predictor, thus at least partially cancelling the initial hypothesis that prediction of difficult queries and personalization performance are highly correlated. Because of the previous correlations variability and the fact that each predictor measures different aspects, we think of interest to join the potential benefit of each of the individual PPP using classification and regression techniques. In fact, those techniques offer an additional advantage, since they directly give us the response to whether or not to apply personalization, in contrast with any individual predictor, for which we would need to select a threshold for that decision problem.

\subsection{Using classification and regression techniques}
\label{subsec:classRegression}

To try to improve the prediction performance joining all the predictors potential by using classification and regression techniques, we have used the WEKA\footnote{\url{http://www.cs.waikato.ac.nz/ml/weka/}} machine learning framework. 

As correlations highly depend on the user profile applied to the query and these profiles have a different number of evaluation triplets (see Table \ref{tab:tripletsByProfile}), we have decided to build a different predictive model for each of the profiles. The input data are the values of all the predictors (which play the role of feature variables) for each evaluation triplet. The last variable (the class) is \textit{diffPerso}. As \textit{diffPerso} is a numeric value it is directly used for regression, but it must be categorized for classification. Our prediction problem is a binary decision: to personalize (\textit{\dquotes{yes}}) or not to personalize (\textit{\dquotes{no}}). We have categorized \textit{diffPerso} in the following way: if $diffPerso < 0$ we transform it into the nominal \textit{\dquotes{no}} (personalization harms the original query performance), if $diffPerso > 0$ we transform it into the nominal \textit{\dquotes{yes}} (personalization improves the original query performance), and if $diffPerso = 0$ we delete this observation, since it does not provide any useful information for personalization (it may be considered as \squotes{noise}).

We have tried several classification and regression algorithms implemented in WEKA, but finally we have decided to use the \textit{Random Forest} approach, since it provides the best results and at the same time it is suitable for both classification and regression.

Personalization almost always outperforms the original query performance. For this reason, the approach followed by almost all personalized IRSs is to always personalize all the queries. This affirmation is validated by our own results, representing the 76 and 82 percent of all the triplets (see again Table \ref{tab:tripletsByProfile}) for the User Study and ASPIRE, respectively. Roughly speaking, we could say that 3 out of 4 queries are helped by personalization. For this reason the baseline approach to compare with is to always personalize.

We have applied a \textit{Leave-One-Out (LOO)} approach for the user study, since it has only a few observations for each profile (see Table \ref{tab:tripletsByProfile}). For ASPIRE we have used a 10-fold cross validation approach to evaluate the prediction model.

Tables \ref{tab:classRegUserStudy} and \ref{tab:classRegAspire} show the results of the learned models. \textit{avgPerso} is the NDCG average over all the test observations always using personalization (\textit{baseline}). The next two columns \textit{avgPred} and \textit{\% gain} are the average over all the test observations and the percentage gain against \textit{avgPerso} using the best performance approach, i.e., this \textit{ideal} value would be obtained if we were able to always make the best decision about using the original or the personalized query. The next four columns are the results following the predictions given by the built classification and regression models. The last row in the tables, \dquotes{\textit{$\mu$ IdealGain \%}} represents the percentage of the ideal \textit{\% gain} we are able to catch following our prediction models.

\begin{table}[htbp]
	\footnotesize
	\centering
	\caption{Leave-One-Out Random Forest classification and regression prediction results for the User Study.}
	\begin{tabular}{|c||c|cc||cc|cc|}
		\hline
		& \textit{(baseline)} & \multicolumn{2}{|c||}{\textbf{Ideal}} & \multicolumn{2}{|c|}{\textbf{Classification}} & \multicolumn{2}{|c|}{\textbf{Regression}} \\
		\textbf{User profile} & \textbf{avgPerso} & \textbf{avgPred} & \% \textbf{gain} & \textbf{avgPred} & \% \textbf{gain} & \textbf{avgPred} & \% \textbf{gain} \\ \hline
		\textbf{administration} & 0.608051 & 0.644634 & 6.01644 & 0.570279 & -6.21198 & 0.592527 & -2.55308 \\
		\textbf{agriculture} & 0.79087 & 0.810896 & 2.53215 & 0.774176 & -2.11084 & 0.774176 & -2.11084 \\
		\textbf{culture} & 0.638095 & 0.718654 & 12.62492 & 0.689769 & 8.09817 & 0.689769 & 8.09817 \\
		\textbf{economy} & 0.40522 & 0.545995 & 34.74039 & 0.474091 & 16.99595 & 0.421873 & 4.10962 \\
		\textbf{education} & 0.563201 & 0.592569 & 5.21448 & 0.563201 & 0 & 0.563201 & 0 \\
		\textbf{employment} & 0.617909 & 0.656129 & 6.18538 & 0.614254 & -0.59151 & 0.627099 & 1.48727 \\
		\textbf{environment} & 0.611982 & 0.687295 & 12.30641 & 0.641862 & 4.8825 & 0.658471 & 7.59647 \\
		\textbf{health} & 0.717045 & 0.722 & 0.7132 & 0.717045 & 0 & 0.717045 & 0 \\ \hline
		$\mu$ & 0.619047 & 0.672291 & 10.04167 & 0.630585 & 2.63279 & 0.63052 & 2.07845 \\
		$\mu$ \textit{IdealGain} \% & -- & -- & -- & -- & 26.22 & -- & 20.70 \\ \hline
	\end{tabular}
	\label{tab:classRegUserStudy}
\end{table}

\begin{table}[htbp]
	\footnotesize
	\centering
	\caption{Random Forest classification and regression prediction results for ASPIRE.}
	\begin{tabular}{|c||c|cc||cc|cc|}
		\hline
		& \textit{(baseline)} & \multicolumn{2}{|c||}{\textbf{Ideal}} & \multicolumn{2}{|c|}{\textbf{Classification}} & \multicolumn{2}{|c|}{\textbf{Regression}} \\
		\textbf{User profile} & \textbf{avgPerso} & \textbf{avgPred} & \% \textbf{gain} & \textbf{avgPred} & \% \textbf{gain} & \textbf{avgPred} & \% \textbf{gain} \\ \hline
		\textbf{administration} & 0.703849 & 0.736331 & 4.61491 & 0.719888 & 2.27876 & 0.716649 & 1.81857 \\
		\textbf{agriculture} & 0.834488 & 0.858497 & 2.87709 & 0.823421 & -1.3262 & 0.846435 & 1.43166 \\
		\textbf{culture} & 0.782663 & 0.80918 & 3.38805 & 0.793442 & 1.37722 & 0.794417 & 1.5018 \\
		\textbf{economy} & 0.688553 & 0.724375 & 5.2025 & 0.699786 & 1.63139 & 0.705563 & 2.4704 \\
		\textbf{education} & 0.841959 & 0.844191 & 0.2651 & 0.84129 & -0.07946 & 0.842851 & 0.10594 \\
		\textbf{employment} & 0.779812 & 0.78025 & 0.05617 & 0.779812 & 0 & 0.779812 & 0 \\
		\textbf{environment} & 0.801694 & 0.804977 & 0.40951 & 0.80154 & -0.01921 & 0.80154 & -0.01921 \\
		\textbf{health} & 0.818913 & 0.82685 & 0.96921 & 0.811543 & -0.89997 & 0.816464 & -0.29905 \\ \hline
		$\mu$ & 0.781491 & 0.798081 & 2.22282 & 0.783840 & 0.37032 & 0.787966 & 0.87626 \\
		$\mu$ \textit{IdealGain} \% & -- & -- & -- & -- & 16.66 & -- & 39.42 \\ \hline		
	\end{tabular}
	\label{tab:classRegAspire}
\end{table}

We can draw the following conclusions from Tables \ref{tab:classRegUserStudy} and \ref{tab:classRegAspire}: 1) all \textit{\% gain} (ideal, classification and regression) are less robust between profiles in the User study than in ASPIRE; 2) the ideal \textit{\% gain} are relatively low for both approaches, being the value for the User study considerably higher than for ASPIRE, probably because of the less robustness of the User study; 3) for the User study the \textit{\% gain} for classification and regression are similar on average although different across profiles. For the User study in the best case (classification) we are able to identify approximately 1 out of 4 triplets where personalization harms the original query performance; 4) for ASPIRE the \textit{\% gain} obtained by the regression model clearly outperforms the value obtained by the classification model. In this case we are able to identify a bit more than 1 out of 3 triplets where personalization harms the original query performance.

We trust more on the ASPIRE results because in the User study there are not enough data to obtain robust results, even using the LOO approach for building the predictive models. Another hint for this lack of robustness is that this problem even being a better fit for regression (since the class is numeric and by categorizing it you lose some information) under the User study classification performs better, unlike ASPIRE. We show the User study results for comparison purposes with ASPIRE and mainly because their relevance assessments were provided by real users. But unfortunately, they do not represent enough data to make generalizations based on them. Other things we have tried in order to improve the ASPIRE results is to resample the observations following different strategies, due to the imbalanced dataset, especially in some profiles. Another attempt was to learn only one general classifier and regressor for all the profiles. Both attempts gave us worse results than those in Table \ref{tab:classRegAspire}.

The problem (depending on how you look at it) is that applying personalization is beneficial for almost all queries. Therefore, it is very difficult to accurately predict those queries not benefited without failing to predict those benefited ones. There is a low maximum gain value from which we are able to reach a bit more than one third. \\

\textit{\textbf{Predictors selection.}} To get the previous prediction performance results we need to calculate all the 37 pre-retrieval predictors proposed in this article. The next step is to check if we are able to reach similar prediction values using a considerable less number of predictors, which obviously will be faster in calculation and response time. We will do this process only for ASPIRE because its results are more robust and trustworthy.

For this task we could follow two different alternatives: to use any of the automatic feature selection strategies available in Weka or to manually select the features with the higher correlations from Table \ref{tab:corsBests}. We have explored some of the strategies from the first alternative including: the \textit{CfsSubsetEval}, which provides a set of predictors by considering the individual predictive ability of each predictor along with the degree of redundancy between them; \textit{CorrelationAttributeEval} and \textit{InfoGainAttributeEval}, which provide a ranked list of predictors by measuring the correlation and the information gain, respectively, between them and the class, and some others with different parameter configurations with no good final results. However, we obtain almost the same prediction performance results than using all the proposed predictors by using the 10 predictors with the highest ASPIRE correlations from Table \ref{tab:corsBests}. Table \ref{tab:classRegFS} shows these new results.

\begin{table}[htbp]
	\footnotesize
	\centering
	\caption{Random Forest classification and regression prediction results for ASPIRE using the 10 highest correlation predictors from Table \ref{tab:corsBests}.}
	\begin{tabular}{|c||c|cc||cc|cc|}
		\hline
		& \textit{(baseline)} & \multicolumn{2}{|c||}{\textbf{Ideal}} & \multicolumn{2}{|c|}{\textbf{Classification}} & \multicolumn{2}{|c|}{\textbf{Regression}} \\
		\textbf{User profile} & \textbf{avgPerso} & \textbf{avgPred} & \% \textbf{gain} & \textbf{avgPred} & \% \textbf{gain} & \textbf{avgPred} & \% \textbf{gain} \\ \hline
		\textbf{administration} & 0.703849 & 0.736331 & 4.61491 & 0.716764 & 1.83491 & 0.722564 & 2.65895 \\
		\textbf{agriculture} & 0.834488 & 0.858497 & 2.87709 & 0.829471 & -0.60121 & 0.829797 & -0.56214 \\
		\textbf{culture} & 0.782663 & 0.80918 & 3.38805 & 0.788394 & 0.73224 & 0.789665 & 0.89464 \\
		\textbf{economy} & 0.688553 & 0.724375 & 5.2025 & 0.698906 & 1.50359 & 0.710254 & 3.15168 \\
		\textbf{education} & 0.841959 & 0.844191 & 0.2651 & 0.84129 & -0.07946 & 0.842145 & 0.02209 \\
		\textbf{employment} & 0.779812 & 0.78025 & 0.05617 & 0.779812 & 0 & 0.779812 & 0 \\
		\textbf{environment} & 0.801694 & 0.804977 & 0.40951 & 0.80154 & -0.01921 & 0.80154 & -0.01921 \\
		\textbf{health} & 0.818913 & 0.82685 & 0.96921 & 0.814721 & -0.5119 & 0.818077 & -0.10209 \\ \hline
		$\mu$ & 0.781491 & 0.798081 & 2.22282 & 0.783862 & 0.35737 & 0.786732 & 0.75549 \\
		$\mu$ \textit{IdealGain} \% & -- & -- & -- & -- & 16.08 & -- & 33.99 \\ \hline		
	\end{tabular}
	\label{tab:classRegFS}
\end{table}

Looking at the results of Table \ref{tab:classRegFS}, we can see how almost exactly the same prediction performance is reached by using classification and a slightly worse performance in the case of regression, although still being able to capture 1 out of 3 triplets where personalization harms the original query performance. If we only use the five predictors with the highest correlation values from Table \ref{tab:corsBests}, the prediction power drops to approximately half than using all the predictors and this is not acceptable.

Therefore, the conclusion is that if the final IRS time response is critical it should only use the ten ASPIRE predictors from Table \ref{tab:corsBests} to decide whether to personalize or not the user query, since almost the same prediction performance is reached than using all the proposed predictors in this article. If the IRS time response is not so critical both approaches could be used, since their difference in time computation is not significant against the time required to perform the search.

\section{Conclusions and Future Work}
\label{sec:conclusionsFuture}

In this article we have faced the difficult task to predict whether personalization will benefit or harm the original query performance before the search is performed. If we are able to identify the harmed queries the personalization module could be deactivated for those cases in order to obtain the maximum performance from the personalized IRS. Most of the times personalization outperforms the original query performance but this is not always true and it will depend both on the query and the user.

Within the literature this personalization prediction problem has been related to the problem of predicting difficult queries. A difficult query is one that have so many possible answers that it is difficult to retrieve the most appropriate ones. This normally happens when the query is very short, ambiguous or its topic is very general. In those situations personalization helps to provide closer results to the user being this one more satisfied with the IRS.

We have performed a comprehensive study using most of the state-of-the-art pre-retrieval query difficulty predictors. These predictors are based on different assumptions on how users assign relevance to the list of retrieved results. As in personalization results not only the query but also the profile has some influence, we have extended the previous predictors and proposed some others to also include the profile information into the personalization prediction problem. We have finally used a comprehensive and heterogeneous set of 37 pre-retrieval predictors.

We have correlated these predictors with \textit{diffPerso}, the difference between the performance of the personalized query and the original query. Since these correlations are not very high, there is no one predictor which could be considered good enough to predict the personalization performance. Consequently, we have tried to get the most of each predictor potential, based on different assumptions, considering all of them together by the use of classification and regression techniques.

As far as we know nobody else has carried out such a comprehensive study, including the use of machine learning techniques, and give the final improvement of their personalization prediction models against the logical baseline of always apply personalization to all queries. Our built personalization prediction models are able to improve the personalization performance a bit more than one third of the maximum reachable ideal performance, i.e., to be able to identify and therefore disable personalization for all the personalized queries with lower performance than their corresponding original query. We also finally prove that by only using the 10 predictors with the highest \textit{diffPerso} correlations, and not all the 37 proposed in the article, almost the same improvement could be reached. This may be important for IRSs where the query response time is critical.

In general, we believe that this article results are promising and a good starting point to continue the research in this area. We think new predictors are needed, some of them probably using new ways to include the user profile information, to improve our results and being as close as possible to the maximum ideal personalization performance. Additionally to propose new personalization performance predictors, another future work could be to include not all the user profile information but the part more related to the given query, especially important if the profiles are heterogeneous representing several areas of interest.

\begin{acknowledgements}
This work has been supported by the Spanish Andalusian \dquotes{Consejer\'ia de Innovaci\'on, Ciencia y Empresa} postdoctoral phase of project P09-TIC-4526, the Spanish \dquotes{Ministerio de Econom\'ia y Competitividad} projects TIN2013-42741-P and TIN2016-77902-C3-2-P, and the European Regional Development Fund (ERDF-FEDER).
\end{acknowledgements}

\appendix

\section{Appendix}
\label{app:appendixA}

\begin{table}[ht]
	\centering
	\caption{\textit{diffPerso}-predictors correlation values by user profile and predictor for the User Study. \textit{Note:} for '--' values \textit{diffPerso} and predictor values are the same for all the evaluation triplets and given profile, therefore the standard deviation is zero and there is no correlation value.}
	\resizebox{\linewidth}{!}{
	\begin{tabular}{c|cccccccc|}
		& \textbf{administration} & \textbf{agriculture} & \textbf{culture} & \textbf{economy} & \textbf{education} & \textbf{employment} & \textbf{environment} & \textbf{health} \\
		\hline
		\textbf{numQT} & 0.036 & -0.201 & -0.035 & 0.005 & 0.087 & -0.308 & -0.015 & -0.572 \\
		\textbf{avgQL} & 0.41  & -0.211 & -0.006 & 0.059 & 0.237 & -0.027 & 0.053 & -0.334 \\
		\textbf{sumIDF} & -0.193 & -0.104 & -0.223 & 0.396 & -0.245 & -0.068 & -0.474 & -0.725 \\
		\textbf{avgIDF} & -0.008 & 0.064 & -0.218 & 0.45  & -0.169 & 0.368 & -0.401 & -0.472 \\
		\textbf{maxIDF} & -0.299 & 0     & -0.446 & 0.391 & -0.229 & 0.082 & -0.554 & -0.673 \\
		\textbf{sumICTF} & 0.02  & -0.01 & -0.052 & 0.418 & -0.162 & -0.011 & -0.263 & -0.619 \\
		\textbf{avgICTF} & 0.14  & 0.25  & 0.008 & 0.572 & -0.198 & 0.526 & -0.299 & -0.316 \\
		\textbf{maxICTF} & -0.237 & 0.15  & -0.471 & 0.464 & -0.242 & 0.237 & -0.428 & -0.483 \\
		\textbf{SCS} & 0.15  & 0.315 & 0.068 & 0.501 & -0.197 & 0.583 & -0.197 & -0.105 \\
		\textbf{sumSCQ} & -0.134 & -0.269 & -0.135 & 0.144 & -0.032 & -0.261 & -0.34 & -0.755 \\
		\textbf{avgSCQ} & -0.061 & -0.044 & -0.146 & 0.355 & -0.099 & 0.359 & -0.463 & -0.433 \\
		\textbf{maxSCQ} & -0.35 & -0.101 & -0.345 & 0.313 & -0.12 & 0.148 & -0.751 & -0.649 \\
		\textbf{sumVAR} & -0.174 & -0.088 & -0.151 & 0.215 & -0.301 & -0.248 & -0.087 & -0.479 \\
		\textbf{avgVAR} & -0.184 & -0.092 & -0.277 & 0.161 & -0.296 & -0.258 & -0.111 & -0.491 \\
		\textbf{maxVAR} & -0.179 & -0.084 & -0.18 & 0.17  & -0.304 & -0.25 & -0.082 & -0.481 \\
		\textbf{joint} & -0.182 & -0.09 & -0.158 & 0.254 & -0.301 & -0.243 & -0.101 & -0.489 \\
		\textbf{joint2} & -0.187 & -0.087 & -0.188 & 0.215 & -0.303 & -0.244 & -0.098 & -0.493 \\
		\textbf{cosineQP} & 0.252 & -0.444 & 0.071 & 0.311 & -- & -0.755 & -0.672 & -0.536 \\
		\textbf{sumIDFQP} & -0.193 & -0.104 & -0.223 & 0.396 & -0.245 & -0.068 & -0.474 & -0.725 \\
		\textbf{avgIDFQP} & -0.209 & -0.029 & -0.222 & 0.369 & -0.305 & 0.056 & -0.557 & -0.652 \\
		\textbf{maxIDFQP} & -0.105 & -0.084 & -0.029 & 0.37  & -0.308 & -0.232 & -0.048 & -0.372 \\
		\textbf{sumICTFQP} & 0.02  & -0.01 & -0.052 & 0.418 & -0.162 & -0.011 & -0.263 & -0.619 \\
		\textbf{avgICTFQP} & 0.007 & 0.121 & -0.042 & 0.534 & -0.288 & 0.204 & -0.413 & -0.542 \\
		\textbf{maxICTFQP} & -0.263 & 0.059 & -0.535 & 0.367 & -0.36 & -0.253 & -0.176 & -0.372 \\
		\textbf{SCSQP} & -0.002 & 0.185 & -0.028 & 0.494 & -0.323 & 0.322 & -0.43 & -0.446 \\
		\textbf{sumSCQQP} & -0.134 & -0.269 & -0.135 & 0.144 & -0.032 & -0.261 & -0.34 & -0.755 \\
		\textbf{avgSCQQP} & -0.302 & -0.11 & -0.169 & 0.223 & -0.243 & 0.1   & -0.681 & -0.565 \\
		\textbf{maxSCQQP} & -0.001 & -0.052 & -- & -- & -- & -- & -- & -- \\
		\textbf{sumVARQP} & 0.005 & -0.189 & -0.116 & -0.232 & 0.217 & -0.492 & 0.361 & -0.051 \\
		\textbf{avgVARQP} & -0.019 & -0.17 & -0.137 & -0.313 & 0.237 & -0.509 & 0.48  & 0.124 \\
		\textbf{maxVARQP} & 0.034 & -0.181 & -0.115 & -0.241 & 0.272 & -0.471 & 0.376 & -0.009 \\
		\textbf{jointQP} & 0.005 & -0.189 & -0.116 & -0.232 & 0.217 & -0.492 & 0.361 & -0.051 \\
		\textbf{joint2QP} & 0.034 & -0.181 & -0.115 & -0.241 & 0.272 & -0.471 & 0.376 & -0.009 \\
		\textbf{profIDF} & 0     & -0.069 & 0.217 & -0.451 & 0.162 & -0.378 & 0.39  & 0.459 \\
		\textbf{profICTF} & -0.145 & -0.256 & -0.011 & -0.57 & 0.192 & -0.537 & 0.29  & 0.301 \\
		\textbf{profSCQ} & 0.051 & 0.041 & 0.144 & -0.359 & 0.094 & -0.366 & 0.451 & 0.424 \\
		\textbf{profVAR} & 0.175 & 0.008 & 0.158 & -0.35 & 0.368 & 0.01  & 0.344 & 0.482 \\
		\hline
	\end{tabular}
	}
	\label{tab:corsUserStudyByProfiles}
\end{table}

\begin{table}[ht]
	\centering
	\caption{\textit{diffPerso}-predictors correlation values by user profile and predictor for ASPIRE.}
	\resizebox{\linewidth}{!}{
	\begin{tabular}{c|cccccccc|}
		& \textbf{administration} & \textbf{agriculture} & \textbf{culture} & \textbf{economy} & \textbf{education} & \textbf{employment} & \textbf{environment} & \textbf{health} \\
		\hline
		\textbf{numQT} & 0.132 & -0.001 & 0.208 & 0.048 & 0.011 & -0.315 & -0.231 & 0.021 \\
		\textbf{avgQL} & -0.112 & -0.047 & -0.111 & -0.185 & -0.203 & 0.178 & -0.106 & -0.157 \\
		\textbf{sumIDF} & 0.164 & -0.013 & 0.129 & 0.062 & 0.038 & -0.096 & -0.159 & -0.144 \\
		\textbf{avgIDF} & 0.003 & -0.072 & -0.074 & -0.016 & -0.014 & 0.184 & -0.069 & -0.162 \\
		\textbf{maxIDF} & 0.108 & 0.04  & 0.105 & 0.08  & 0.084 & 0.054 & -0.058 & 0.017 \\
		\textbf{sumICTF} & 0.181 & 0.012 & 0.185 & 0.072 & 0.043 & -0.229 & -0.189 & -0.089 \\
		\textbf{avgICTF} & 0.045 & -0.001 & -0.02 & 0.046 & 0.022 & 0.158 & -0.002 & -0.186 \\
		\textbf{maxICTF} & 0.141 & 0.08  & 0.109 & 0.111 & 0.138 & 0.056 & -0.034 & -0.009 \\
		\textbf{SCS} & -0.069 & -0.043 & -0.139 & -0.022 & -0.04 & 0.292 & 0.073 & -0.192 \\
		\textbf{sumSCQ} & 0.154 & -0.022 & 0.186 & 0.047 & -0.003 & -0.304 & -0.251 & -0.054 \\
		\textbf{avgSCQ} & -0.041 & -0.165 & -0.181 & -0.099 & -0.105 & 0.087 & -0.127 & -0.206 \\
		\textbf{maxSCQ} & 0.069 & 0.005 & 0.028 & 0.039 & -0.088 & -0.02 & -0.142 & 0.011 \\
		\textbf{sumVAR} & 0.02  & -0.057 & 0.137 & -0.026 & 0.043 & -0.017 & 0.038 & 0.057 \\
		\textbf{avgVAR} & 0.047 & -0.051 & 0.141 & -0.014 & 0.084 & 0.066 & 0.102 & 0.125 \\
		\textbf{maxVAR} & 0.018 & -0.057 & 0.137 & -0.025 & 0.05  & -0.016 & 0.057 & 0.072 \\
		\textbf{joint} & 0.02  & -0.057 & 0.137 & -0.026 & 0.043 & -0.017 & 0.038 & 0.057 \\
		\textbf{joint2} & 0.018 & -0.057 & 0.137 & -0.025 & 0.05  & -0.016 & 0.056 & 0.072 \\
		\textbf{cosineQP} & -0.28 & -0.299 & -0.259 & -0.346 & -0.376 & -0.329 & -0.203 & -0.141 \\
		\textbf{sumIDFQP} & 0.164 & -0.013 & 0.129 & 0.062 & 0.038 & -0.096 & -0.159 & -0.144 \\
		\textbf{avgIDFQP} & 0.11  & -0.016 & -0.044 & 0.045 & 0.034 & 0.158 & 0.023 & -0.173 \\
		\textbf{maxIDFQP} & 0.102 & 0.042 & 0.121 & 0.074 & 0.122 & 0.089 & -0.007 & 0.029 \\
		\textbf{sumICTFQP} & 0.181 & 0.012 & 0.185 & 0.072 & 0.043 & -0.229 & -0.189 & -0.089 \\
		\textbf{avgICTFQP} & 0.157 & 0.037 & 0.008 & 0.087 & 0.066 & 0.115 & 0.058 & -0.189 \\
		\textbf{maxICTFQP} & 0.138 & 0.064 & 0.101 & 0.078 & 0.146 & 0.158 & -0.009 & -0.024 \\
		\textbf{SCSQP} & 0.051 & 0.026 & -0.108 & 0.026 & 0.043 & 0.288 & 0.17  & -0.165 \\
		\textbf{sumSCQQP} & 0.154 & -0.022 & 0.186 & 0.047 & -0.003 & -0.304 & -0.251 & -0.054 \\
		\textbf{avgSCQQP} & 0.051 & -0.094 & -0.169 & -0.016 & -0.065 & 0.139 & 0.009 & -0.201 \\
		\textbf{maxSCQQP} & 0.056 & 0.048 & 0.039 & 0.059 & -0.025 & 0.01  & -0.006 & 0.052 \\
		\textbf{sumVARQP} & 0.025 & -0.056 & 0.187 & -0.023 & 0.04  & -0.109 & -0.018 & 0.08 \\
		\textbf{avgVARQP} & 0.029 & -0.054 & 0.185 & -0.023 & 0.057 & -0.064 & 0.018 & 0.095 \\
		\textbf{maxVARQP} & 0.017 & -0.058 & 0.137 & -0.027 & 0.047 & -0.039 & 0.055 & 0.077 \\
		\textbf{jointQP} & 0.025 & -0.056 & 0.187 & -0.023 & 0.04  & -0.109 & -0.018 & 0.08 \\
		\textbf{joint2QP} & 0.017 & -0.058 & 0.137 & -0.027 & 0.047 & -0.039 & 0.055 & 0.077 \\
		\textbf{profIDF} & 0.007 & 0.076 & 0.076 & 0.025 & 0.019 & -0.185 & 0.078 & 0.159 \\
		\textbf{profICTF} & -0.032 & 0.006 & 0.023 & -0.039 & -0.017 & -0.161 & 0.009 & 0.184 \\
		\textbf{profSCQ} & 0.049 & 0.17  & 0.179 & 0.108 & 0.108 & -0.08 & 0.141 & 0.203 \\
		\textbf{profVAR} & -0.05 & 0.048 & -0.132 & 0.011 & -0.087 & -0.086 & -0.118 & -0.128 \\
		\hline
	\end{tabular}
	}
	\label{tab:corsAspireByProfiles}
\end{table}

5\bibliographystyle{spmpsci}		

\end{document}